\documentclass[12pt,preprint]{aastex}

\shorttitle{The Nature of RP Waves Revealed}
\shortauthors{D. S. Bloomfield et~al.}

\begin{document}

\title{The Nature of Running Penumbral Waves Revealed}

\author{D. Shaun Bloomfield, Andreas Lagg, and Sami K. Solanki}
\email{bloomfield@mps.mpg.de}
\affil{Max-Planck-Institut f\"{u}r Sonnensystemforschung, 
	Max-Planck-Str. 2, 37191 Katlenburg-Lindau, Germany}

\begin{abstract}
We seek to clarify the nature of running penumbral (RP) waves: are they 
chromospheric trans-sunspot waves or a visual pattern of upward-propagating 
waves? Full Stokes spectropolarimetric time series of the photospheric 
Si\,{\sc{i}}\,10827~\AA\ line and the chromospheric He\,{\sc{i}}\,10830~\AA\ 
multiplet were inverted using a Milne-Eddington atmosphere. Spatial pixels 
were paired between the outer umbral/inner penumbral photosphere and the 
penumbral chromosphere using inclinations retrieved by the inversion and the 
dual-height pairings of line-of-sight velocity time series were studied for 
signatures of wave propagation using a Fourier phase difference analysis. The 
dispersion relation for radiatively cooling acoustic waves, modified to 
incorporate an inclined propagation direction, fits well the observed phase 
differences between the pairs of photospheric and chromospheric pixels. We 
have thus demonstrated that RP waves are in effect low-$\beta$ slow-mode waves 
propagating along the magnetic field.
\end{abstract}


\section{INTRODUCTION}
\label{sec:int}
The term running penumbral (RP) wave was created by 
\citet{1972ApJ...178L..85Z} to describe chromospheric H$\alpha$ velocity and 
intensity fronts that were observed moving out through sunspot 
penumbrae. Since then, a host of work has been carried out on reporting their 
properties (see, e.g., the series of papers by \citealp{2000A&A...354..305C, 
2001A&A...375..617C} and \citealp{2000A&A...363..306G}), while their exact 
nature has remained unidentified. Currently, the two most likely possibilities 
for the form of these oscillatory disturbances are:
\begin{enumerate}
\item trans-sunspot waves generated in the umbra (e.g., by umbral flashes) and 
limited to the chromospheric layer, 
\item a ``visual pattern'' resulting from field-aligned waves propagating up 
from the photosphere.
\end{enumerate}
To date, many findings point toward RP waves being due to the ``visual 
pattern'' scenario \citep[for an extensive discussion of this topic see the 
recent review by][]{2006RSPTA.364..313B}. This seems especially likely now 
that \citet{2006ApJ...640.1153C} have successfully identified chromospheric 
3-min umbral oscillations as propagating, field-aligned, acoustic waves. 
However, recent work by \citet{2006A&A...456..689T, 2007A&A...463.1153T} has 
not been able to decide between either of the two possible RP wave scenarios. 
Thus, the sum of the evidence is still not conclusive.

In this paper we use velocity time series observations that possess a two-fold 
advantage over those of previous studies. The first is the simultaneous 
recording of photospheric and chromospheric lines, allowing the connection 
between velocities in the lower and upper atmosphere to be accurately 
investigated. The second is the retrieval of the magnetic vector at both 
heights in the atmosphere through the use of full-Stokes spectropolarimetry, 
thus circumventing the need for any assumptions about the possible orientation 
of the magnetic field.

\section{OBSERVATIONAL DATA}
\label{sec:data}
The dataset used here was obtained from the main spot of active region NOAA 
9451 on 2001 May 9 with the Tenerife Infrared Polarimeter 
\citep[{\sc{tip}};][]{1999ASPC..183..264M} attached to the German Vacuum Tower 
Telescope in Tenerife, Canary Islands. The 0.5\arcsec$\times$40\arcsec\ 
spectrograph slit was positioned across NOAA 9451 for approximately 70~min 
with no spatial scanning of the slit, while the solar image was kept 
stationary beneath the slit via a correlation tracking device 
\citep{1996A&AS..115..353B}. The main umbra from this dataset was previously 
analyzed and presented as dataset 2 in the work of 
\citet{2006ApJ...640.1153C}, so only a brief description of the observational 
setup and format of the data is supplied here. Note the correct heliographic 
coordinates of the observed sunspot are S22\degr E20\degr. 

The spectral region obtained in these observations was recorded with a 
wavelength sampling of 31~m\AA~pixel$^{-1}$ and includes the photospheric 
Si\,{\sc{i}}\,10827.09~\AA\ line, the upper-chromospheric 
He\,{\sc{i}}\,10830~\AA\ multiplet (with triplet components at 10829.09~\AA, 
10830.25~\AA, and 10830.34~\AA), and the telluric H$_2$O line at 10832.11~\AA. 
The {\sc{tip}} instrument was used to record simultaneous spectral images of 
the four Stokes parameters ($I$, $Q$, $U$, $V$) for each of the 0.4\arcsec\ 
spatial pixels along the slit. Multiple images were coadded online to increase 
the signal-to-noise ratio, resulting in a final cadence of 2.1~s.

After dark current subtraction, flat-field correction, polarization 
calibration, and removal of polarization cross talk 
\citep{2003SPIE.4843...55C}, Stokes ($I$, $Q$, $U$, $V$) profiles were 
inverted separately for each line using the Milne-Eddington inversion code of 
\citet{2004A&A...414.1109L}. For the He\,{\sc{i}} inversions an atmospheric 
model with one magnetic component was used, while a non-magnetic component was 
included in the Si\,{\sc{i}} inversions to account for stray light. A 
continuum intensity space-time plot of the time series is given in 
Fig.~\ref{fig:con_B_vel}, alongside absolute magnetic field strengths, 
line-of-sight (LOS) velocities, and magnetic field inclinations in solar 
coordinates (see \S~\ref{subsubsec:atm_cou} for more details) retrieved by 
the Si\,{\sc{i}} and He\,{\sc{i}} inversions.

The Si\,{\sc{i}} and He\,{\sc{i}} LOS velocities retrieved from the inversions 
both show systematically decreasing velocity (i.e., blueshift) with time. This 
trend arises from the relative motion along the observer's LOS caused by the 
Earth's rotation, as these observations were obtained during local morning. 
Simple linear fits proved a suitable approximation to the trends and LOS 
velocity time series from each spatial pixel had these linear background 
trends removed prior to any form of temporal analysis.

\section{ANALYSIS METHOD}
\label{sec:ana_met}
Fourier phase difference analysis is a useful tool through which the 
propagation characteristics of waves may be determined. The form of analysis 
used here is based on the standard Fourier equations that are discussed in 
depth by \citet{2001A&A...379.1052K}. This technique has been used extensively 
in the past for various solar studies (e.g., \citealp{1963ApJ...138..252J}, 
\citealp{1973SoPh...33..333S}, \citealp{1984ApJ...277..874L}, 
\citealp{2000ApJ...531.1150W}) and it remains one of the most robust 
methodologies in use.

The phase difference spectrum, $\Delta \phi (\nu)$, between two temporal 
signals measures the phase lag at discrete frequencies. For signals separated 
by some spatial distance, this lag is the cycle-time that it takes different 
frequency components to travel from the first location to that of the second. 
When waves propagate between two locations in a normally dispersive medium, 
phase difference spectra will show zero phase difference at low frequencies 
(where waves are evanescent and not propagating) followed by phase differences 
which increase in magnitude at higher frequencies (as high-frequency 
components travel more slowly than low-frequency components).

The Fourier phase coherence spectrum, $C^2 (\nu)$, between two signals is a 
measure of the quality of phase difference variation. However, unless 
averaging in frequency is performed, the coherence between two Fourier 
components will be unity irrespective of the phase differences. The Fourier 
squared coherence of randomly distributed phase differences then approaches 
1/$n$ for averaging over $n$ points in frequency. In this work coherence 
values are calculated using an average over 5 frequency intervals, providing a 
``noise'' level of 0.2 for randomly distributed (i.e., uncorrelated) phase 
differences. In the following phase difference diagrams, Fourier squared 
coherence is represented by the degree of symbol shading (white for 0; black 
for 1) while the symbol size represents the cross-spectral power -- a measure 
of the co-variance between the Si\,{\sc{i}} and He\,{\sc{i}} LOS velocity 
signals.

\subsection{Trans-sunspot Wave}
\label{subsec:tra_sun_wav}
If RP waves are due to waves propagating across the sunspot chromosphere, 
spectra calculated between the chromospheric LOS velocities from the umbra and 
those from pixels at sequentially greater distances into the penumbra should 
show phase difference values increasing linearly with frequency, becoming 
steeper with greater spatial separation of the signals. In this analysis, 
Fourier phase differences and coherences were calculated between the 
He\,{\sc{i}} LOS velocity from pixel 55 (located in the umbra) and the 
He\,{\sc{i}} LOS velocities from pixels at increasing distances into the 
penumbra.

\begin{figure}
\centering
\includegraphics[width=\columnwidth]{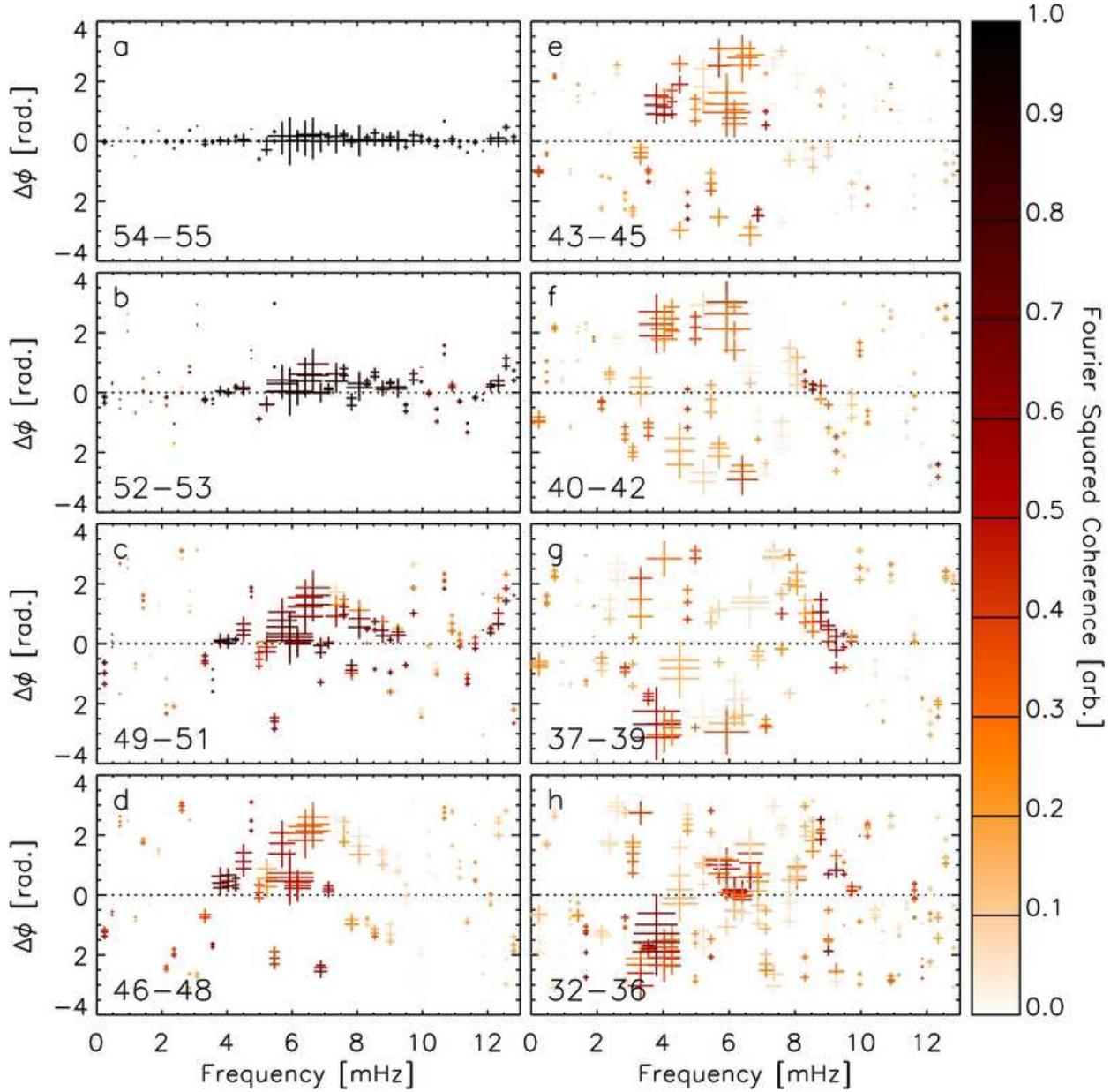}
\caption{Phase difference spectra between He\,{\sc{i}} LOS velocity in the 
umbra and those at increasing distance into the penumbra. Symbol size and 
shading denote cross-spectral power and squared coherence, respectively. 
Panels show groups of phase difference spectra calculated between umbral pixel 
55 and the pixel numbers listed in the lower-left corners, moving from cases 
concerning pixels closest to the umbra/penumbra boundary ({\emph{a}}) toward 
those in the middle penumbra ({\emph{h}}). The Fourier squared coherence 
``noise'' level has a value of 0.2 for randomly distributed phase differences 
in these data.}
\label{fig:fft_tra_sun}
\end{figure}

Figure~\ref{fig:fft_tra_sun} shows the output from such an analysis, where 
phase difference spectra between the He\,{\sc{i}} LOS velocity from pixel 55 
and those pixel numbers listed in each panel are overplotted. For example, 
panel {\emph{a}} contains $\Delta \phi_{\mathrm{He}(t, 
55)\rightarrow\mathrm{He}(t, 55)}$ and $\Delta \phi_{\mathrm{He}(t, 
55)\rightarrow\mathrm{He}(t, 54)}$, while panel {\emph{b}} contains $\Delta 
\phi_{\mathrm{He}(t, 55)\rightarrow\mathrm{He}(t, 53)}$ and 
$\Delta \phi_{\mathrm{He}(t, 55)\rightarrow\mathrm{He}(t, 52)}$, where 
$\mathrm{He}(t, Y)$ denotes the temporal He\,{\sc{i}} LOS velocity signal from 
spatial pixel $Y$. Although groups of phase difference spectra are overplotted 
to increase the clarity of any relations, no clear form of propagating wave 
behaviour is seen. Note that around 4~mHz (i.e., 4-min period) values of 
increasing phase difference are observed when moving into the penumbra. 
However, the finding is marginal since the values of Fourier squared coherence 
in the spectra rapidly approach the ``noise'' level of 0.2 for randomized 
phase differences.

\subsection{Upward-propagating Waves}
\label{subsec:upw_pro_wav}
\begin{figure*}
\centering
\includegraphics[width=15.6cm]{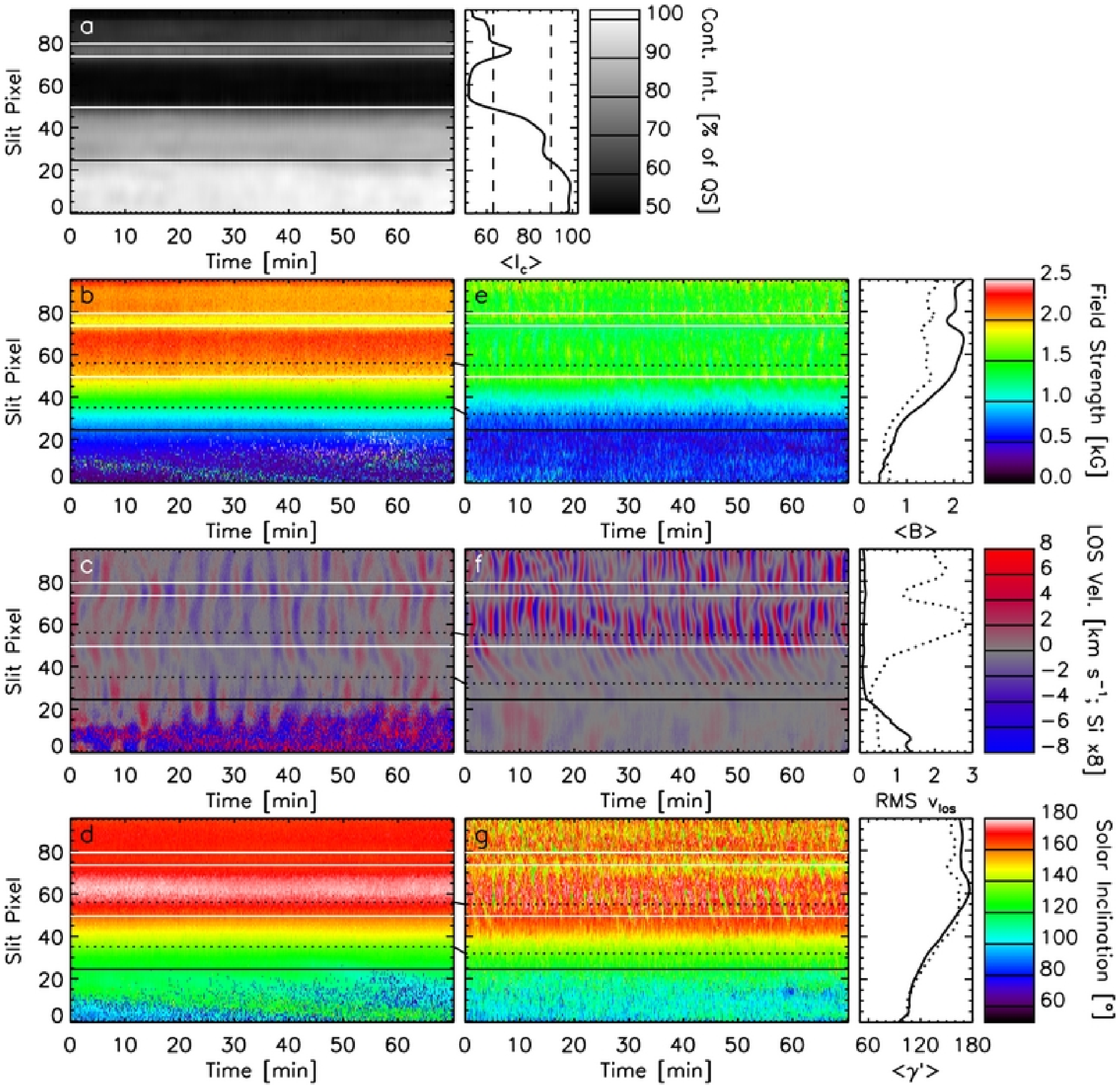}
\caption{Space-time plots of: continuum intensity at 10825.7$\pm$0.3~\AA\ 
({\emph{a}}), absolute magnetic field strengths in Si\,{\sc{i}} ({\emph{b}}) 
and He\,{\sc{i}} ({\emph{e}}), LOS velocities in Si\,{\sc{i}} ({\emph{c}}) and 
He\,{\sc{i}} ({\emph{f}}), and magnetic inclinations in solar coordinates from 
Si\,{\sc{i}} ({\emph{d}}) and He\,{\sc{i}} ({\emph{g}}). Si\,{\sc{i}} and 
He\,{\sc{i}} velocities both have linear background trends removed; 
Si\,{\sc{i}} velocities are scaled up by a factor of 8 to the dynamic range of 
the He\,{\sc{i}} velocities. Upper solid white lines enclose a lightbridge in 
the umbra, while the lower white (black) line marks the umbral/penumbral 
(penumbral/quiet Sun) boundary. Temporal averages of the parameters (RMS 
values for LOS velocities) are shown in the right-most panels for both 
Si\,{\sc{i}} and He\,{\sc{i}} (solid and dotted curves, respectively). Regions 
of spatial pixels paired between Si\,{\sc{i}} and He\,{\sc{i}} (see 
\S~\ref{subsubsec:atm_cou}) are marked by horizontal dotted lines in panels 
{\emph{b}}-{\emph{g}}. Note oscillations in He\,{\sc{i}} field strength and 
inclination are not real, but result from misfitting of the Stokes profiles 
associated with wave shocking. Observed shock profiles require two components; 
the single component inversion used here retrieves weakened, more inclined 
fields. However, the He\,{\sc{i}} velocities retrieved still represent the 
general plasma motion.}
\label{fig:con_B_vel}
\end{figure*}

Although a number of differing forms of propagating wave can exist in the 
outer atmosphere of a sunspot (e.g., fast/slow magneto-acoustic and Alfv\'{e}n 
waves) we shall restrict our analysis of upward-propagating waves to that of 
field-aligned acoustic waves, as these were shown by 
\citet{2006ApJ...640.1153C} to describe the phase behaviour of 3-min waves in 
sunspot umbrae. The extension of these waves from travelling along 
near-vertical field lines in the umbra to travelling along inclined field 
lines which expand out over the penumbra is not unexpected since:
\begin{enumerate}
\item the magnetic field inclination increases smoothly from the umbral centre 
out through the penumbra (Figs.~\ref{fig:con_B_vel}{\emph{d}} and 
\ref{fig:con_B_vel}{\emph{g}}),
\item the photospheric LOS velocity signals are fairly coherent across the 
umbra/penumbra boundary (Fig.~\ref{fig:con_B_vel}{\emph{c}}).
\end{enumerate}

\begin{figure}
\centering
\includegraphics[width=\columnwidth]{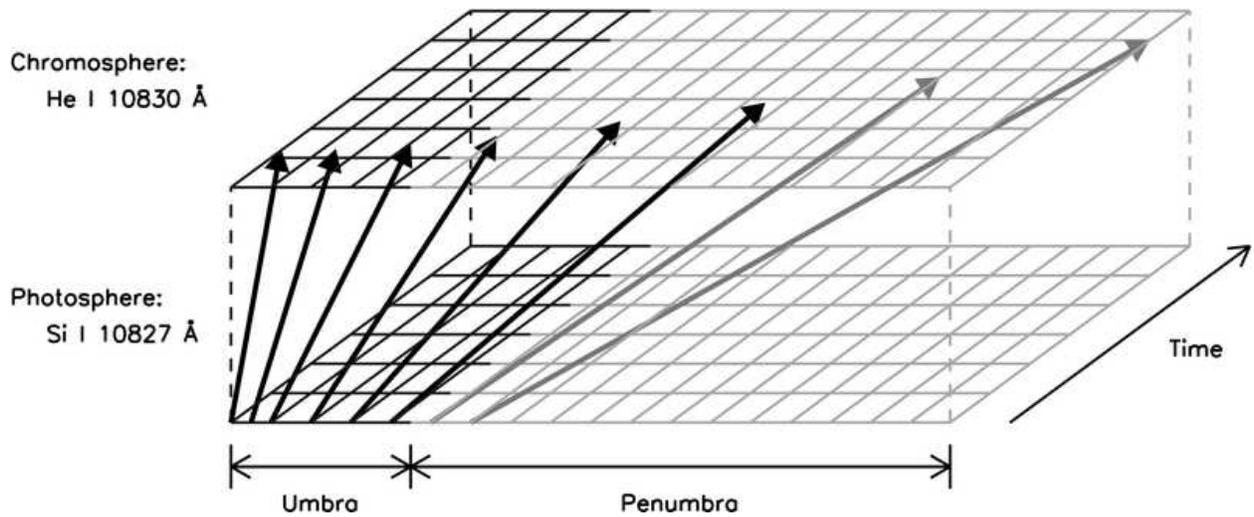}
\caption{Cartoon schematic space-time diagram illustrating the form of pixel 
coupling between the photosphere and chromosphere for the case of 
field-aligned, upward-propagating waves presented in 
\S~\ref{subsec:upw_pro_wav}. Dark (light) grids denote umbral (penumbral) 
pixels, while dark (light) arrows indicate magnetic lines of force (i.e., wave 
paths) linking back to the photospheric umbra (penumbra). Note increasing 
delays in wavefront arrival time at the chromospheric sampling height because 
of increased propagation lengths along more inclined field lines. The 
horizontal and vertical axes are not to scale, resulting in magnetic field 
inclinations that appear different to the values actually retrieved from the 
Stokes inversions.}
\label{fig:cartoon}
\end{figure}

To study the possible propagation of field-aligned waves between two 
atmospheric heights we must first accurately determine the photospheric pixels 
that provide the lower atmospheric signal for the chromospheric pixels which 
lie above the penumbra. This is necessary because the field is significantly 
inclined here (Figs.~\ref{fig:con_B_vel}{\emph{d}} and 
\ref{fig:con_B_vel}{\emph{g}}) and velocity signals in the upper atmosphere 
will be spatially removed from their originating photospheric pixels. The 
expected picture for field-aligned, upward-propagating waves is indicated in 
the schematic diagram of Fig.~\ref{fig:cartoon}, where increasingly inclined 
field lines at the photosphere reach further into the chromospheric penumbra.

\subsubsection{Atmospheric Height Coupling}
\label{subsubsec:atm_cou}
In order to correctly pair spatial pixels between the photosphere and the 
chromosphere we require reliable determination of the magnetic field vector 
in solar coordinates. This is complicated by the 180\degr\ azimuthal 
ambiguity, whereby two equally valid but opposite azimuth orientations exist 
in the observer's coordinate frame. This uncertainty in the field 
azimuth impacts on the whole magnetic vector; the two differing azimuthal 
solutions yield different solar inclinations, $\gamma^{\prime}$.

To overcome this ambiguity we have implemented a ``smoothest magnetic vector'' 
form of ambiguity solution. Namely, a pixel region with the most realistic 
solution is selected as a trusted starting point (e.g., in the umbra where 
the true solution should be closest to vertical). Moving away from this seed 
region, either the 0\degr\ or 180\degr\ azimuth solution is chosen on a 
pixel-by-pixel basis to minimise the spatial variation in the three 
orthogonal components of the solar magnetic vector ($B_x$, $B_y$, $B_z$), 
where the $z$-direction is normal to the solar surface. The field inclinations 
achieved in solar coordinates are shown in Figs.~\ref{fig:con_B_vel}{\emph{d}} 
and \ref{fig:con_B_vel}{\emph{g}} for the inversion results from Si\,{\sc{i}} 
and He\,{\sc{i}}, respectively. 

Spatial pixels were then paired between the photosphere and chromosphere using 
the temporal averages of Si\,{\sc{i}} inclinations in the solar coordinate 
frame\footnote{Inclinations determined by the Si\,{\sc{i}} inversion were 
chosen for this task because the greater signal-to-noise achieved in this line 
means that its magnetic vector, and hence solar inclination, is more reliably 
determined in comparison to that from the He\,{\sc{i}} inversion.}, $\langle 
\gamma^{\prime}_{\mathrm{Si}} \rangle$. Coupled with an expected height 
separation, $\Delta H$, these inclinations provide pixel offsets between 
photospheric and chromospheric pixel pairs in the direction along the slit by 
$\Delta S = \vert \tan \langle \gamma^{\prime}_{\mathrm{Si}} \rangle \vert 
\cos \alpha \Delta H / s_{\mathrm{pix}}$, where $\Delta H$ was taken as 
1000~km \citep[following the findings of][for this sunspot 
umbra]{2006ApJ...640.1153C}, $s_{\mathrm{pix}}$ is the spatial sampling of the 
slit ($\approx$$300$~km\,pixel$^{-1}$), and $\alpha$ is the angle between the 
field azimuth and the slit direction. Regions of pixels paired together in 
this work are outlined in panels {\emph{b}}-{\emph{g}} of 
Fig.~\ref{fig:con_B_vel} by dotted lines, while details of the pixel pairs are 
provided in Table~\ref{tab:pixel_pairs} with their corresponding field 
inclinations. Although this approach uses the simplifying assumption that the 
magnetic field remains essentially linear between the two formation heights 
(i.e., there is no field curvature), it is somewhat justified by the resulting 
pixel pairs in Table~\ref{tab:pixel_pairs} having inclinations that differ by 
$\leqslant$7\degr\ -- i.e., $\langle \gamma^{\prime}_{\mathrm{Si}} \rangle 
\approx \langle \gamma^{\prime}_{\mathrm{He}} \rangle$.

\begin{figure}
\centering
\includegraphics[width=\columnwidth]{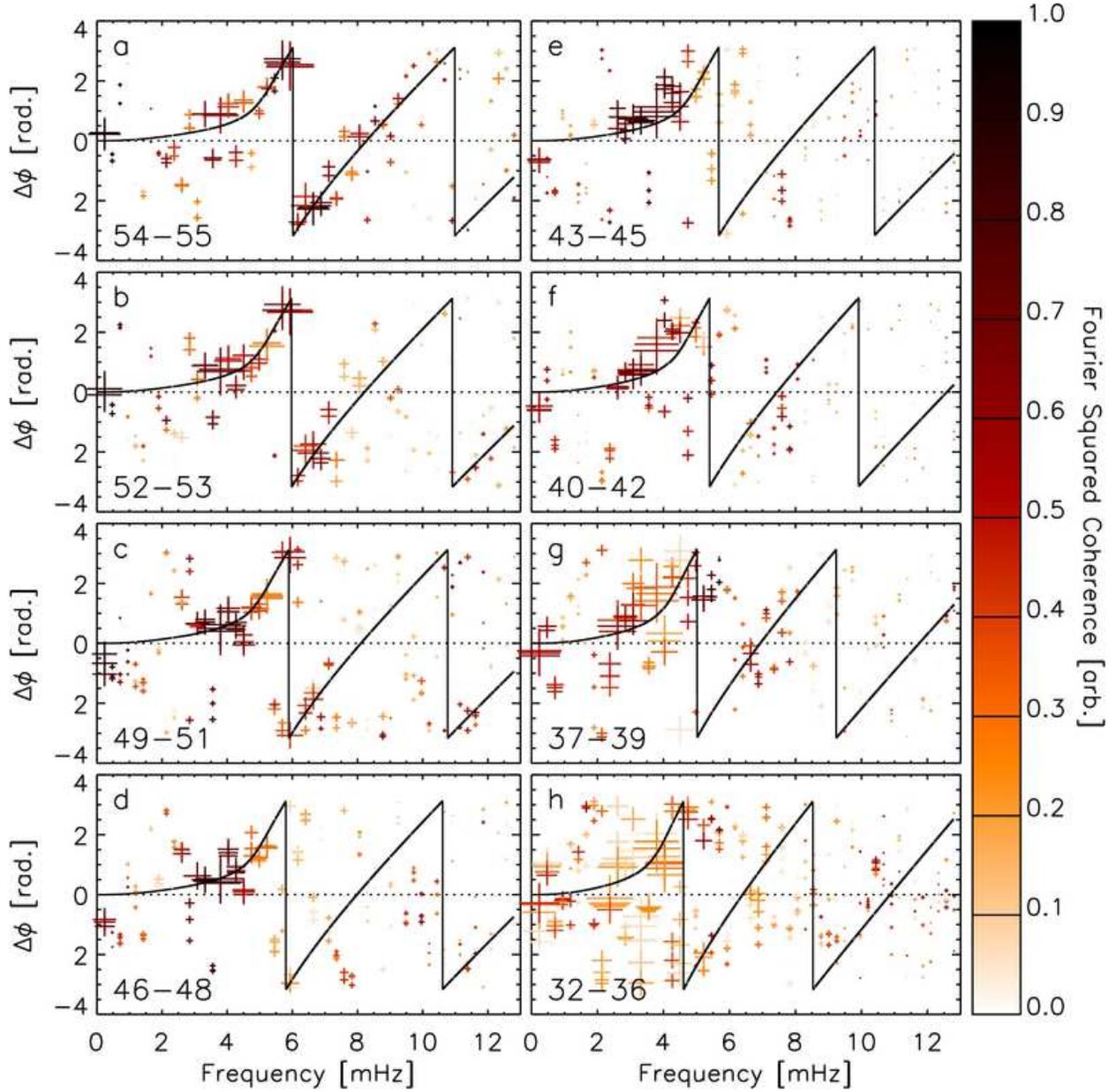}
\caption{Same as Fig.~\ref{fig:fft_tra_sun}, but for spatially-offset 
dual-height pairs of photospheric and chromospheric pixels. Curves show 
modified acoustic dispersion curves using the measured Si\,{\sc{i}} field 
inclinations.}
\label{fig:fft_upw_pro}
\end{figure}

The Fourier phase difference spectra resulting from these dual-height pixel 
pairs are presented in Fig.~\ref{fig:fft_upw_pro}, where spectra from groups 
of adjacent pixel pairs are again overplotted to enhance any relations. In 
contrast to the trans-sunspot case depicted in Fig.~\ref{fig:fft_tra_sun}, the 
expected form of phase difference variation due to propagation (i.e., 
increasing values of phase difference with frequency) is clearly apparent in 
most of the panels. In addition, throughout panels {\emph{a}}-{\emph{g}} 
Fourier squared coherence values remain reasonably high.

\subsubsection{Dispersion Relation Comparison}
\label{subsubsec:dis_rel_fit}
We make use of the equations provided in \citet{2006ApJ...640.1153C} which 
describe the dispersion relation for vertical acoustic waves propagating 
in the presence of a vertical magnetic field within a stratified isothermal 
atmosphere with radiative cooling. The equations were modified for this work 
to simplistically mimic the first order effects that acoustic-like 
(low-$\beta$ slow-mode) waves would experience when propagating along inclined 
field lines instead of purely vertically in a vertical magnetic field -- i.e., 
$\cos \gamma^{\prime}$ reduced gravity and 1/$\cos \gamma^{\prime}$ increased 
path length.

The solid curves in Fig.~\ref{fig:fft_upw_pro} were calculated using the 
measured values of field inclination in solar coordinates along with the 
temperature (4000~K), radiative cooling time (55~s), and vertical height 
separation (1000~km) given by \citet{2006ApJ...640.1153C} for this sunspot 
umbra. Although only a simple approximation to the expected dispersion 
relation for such waves, these curves show an encouraging association with the 
measured data points.

\section{DISCUSSION}
\label{sec:dis}
In this section we present our findings in the context of results from 
previous studies in an attempt to provide answers to a few of the outstanding 
issues surrounding the relationship that RP waves share with other forms of 
sunspot waves.

\begin{figure}
\centering
\includegraphics[width=\columnwidth]{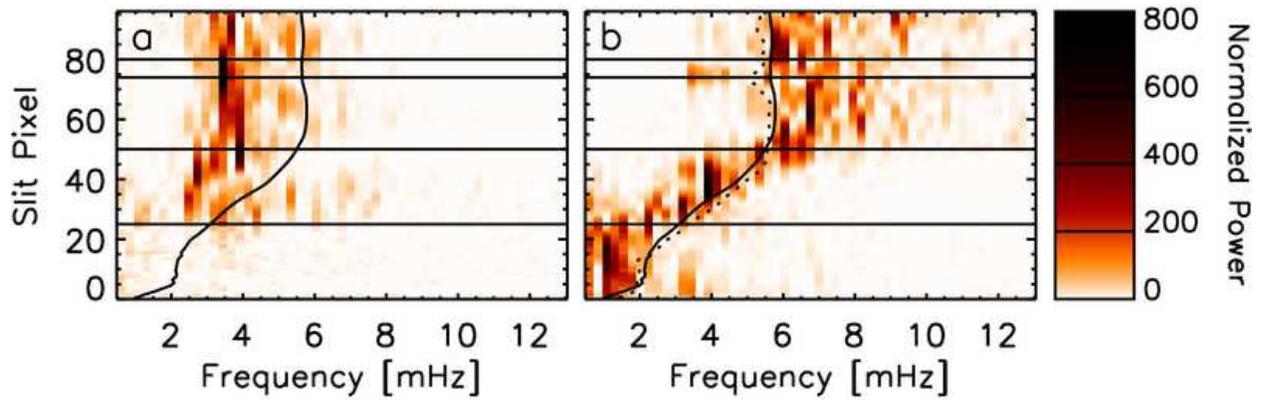}
\caption{Space-frequency plots of Si\,{\sc{i}} ({\emph{a}}) and He\,{\sc{i}} 
({\emph{b}}) velocity power. Spectra from each spatial pixel have been 
normalized to the variance of the corresponding time series; white noise has 
power of unity and 18.4 is the 99.99\% significance level. Overlaid solid 
(dotted) curves are the acoustic cutoff modified by the Si\,{\sc{i}} 
(He\,{\sc{i}}) inclinations, while horizontal lines mark the same boundaries 
as in Fig.~\ref{fig:con_B_vel}.}
\label{fig:he_fft_pow}
\end{figure}

It has been long known that sunspot chromospheres oscillate at differing 
periods in different spatial regions \citep{1972SoPh...27...71G}. 
Figure~\ref{fig:he_fft_pow} displays the variation of Fourier power from the 
Si\,{\sc{i}} and He\,{\sc{i}} LOS velocities in the form of space-frequency 
diagrams. Individual power spectra from each spatial pixel have been 
normalized to the variance of the respective time series, resulting in white 
noise having power of 1 and 18.4 being the 99.99\% significance level of 
Poisson noise. Normalization was performed to aid in the comparison of 
spectral profiles between spatial regions that exhibit vastly different LOS 
velocity amplitudes. Rather than indicating some form of physical 
discontinuity \citep[c.f.,][]{2006A&A...456..689T, 2007A&A...463.1153T}, the 
change from dominant chromospheric 3-min power to longer periods near the 
umbra/penumbra boundary in Fig.~\ref{fig:he_fft_pow}{\emph{b}} may just result 
from the magnetic field inclination becoming large enough to allow 
photospheric low-frequency (i.e., 5-min $p$-mode) power to tunnel through the 
higher-frequency acoustic cutoff (5.2~mHz) at the temperature minimum 
\citep{2004Natur.430..536D}. We note that the classical interpretation of an 
acoustic cutoff is effectively negated by the inclusion of radiative cooling 
in the modeled dispersion relation of \S~\ref{subsubsec:dis_rel_fit}, which 
allows wave reflection and transmission at all frequencies. However, the 
dominant chromospheric frequency in Fig.~\ref{fig:he_fft_pow}{\emph{b}} is 
modified by the magnetic field inclination, closely following the strong-field 
limit $\cos \gamma^{\prime}$ relation of \citet{1977A&A....55..239B}. Power 
existing below the cutoff may be explained by the gradual transition from 
mainly reflected to transmitted waves around the cutoff (i.e., the slow 
turn-up in the curves of Fig.~\ref{fig:fft_upw_pro}). Another possibility is 
the presence of unresolved structure in the chromospheric penumbra 
\citep{1995A&A...293..252R}, consisting of either two spatially-separated 
distributions of field inclination along the LOS or an uncombed magnetic field 
configuration \citep{2007A&A...462.1147L}. If the more vertical distribution 
has the measured field inclinations and the other has values $\sim$20\degr\ 
larger, the acoustic cutoff curve in Fig.~\ref{fig:he_fft_pow}{\emph{b}} could 
be pulled to even lower frequencies. At larger field inclinations (i.e., 
further into the penumbra) little evidence is found of 3-min waves because 
power at 5-min period vastly exceeds that at 3-min in the underlying 
photosphere (Fig.~\ref{fig:he_fft_pow}{\emph{a}}).

\begin{figure}
\centering
\includegraphics[width=\columnwidth]{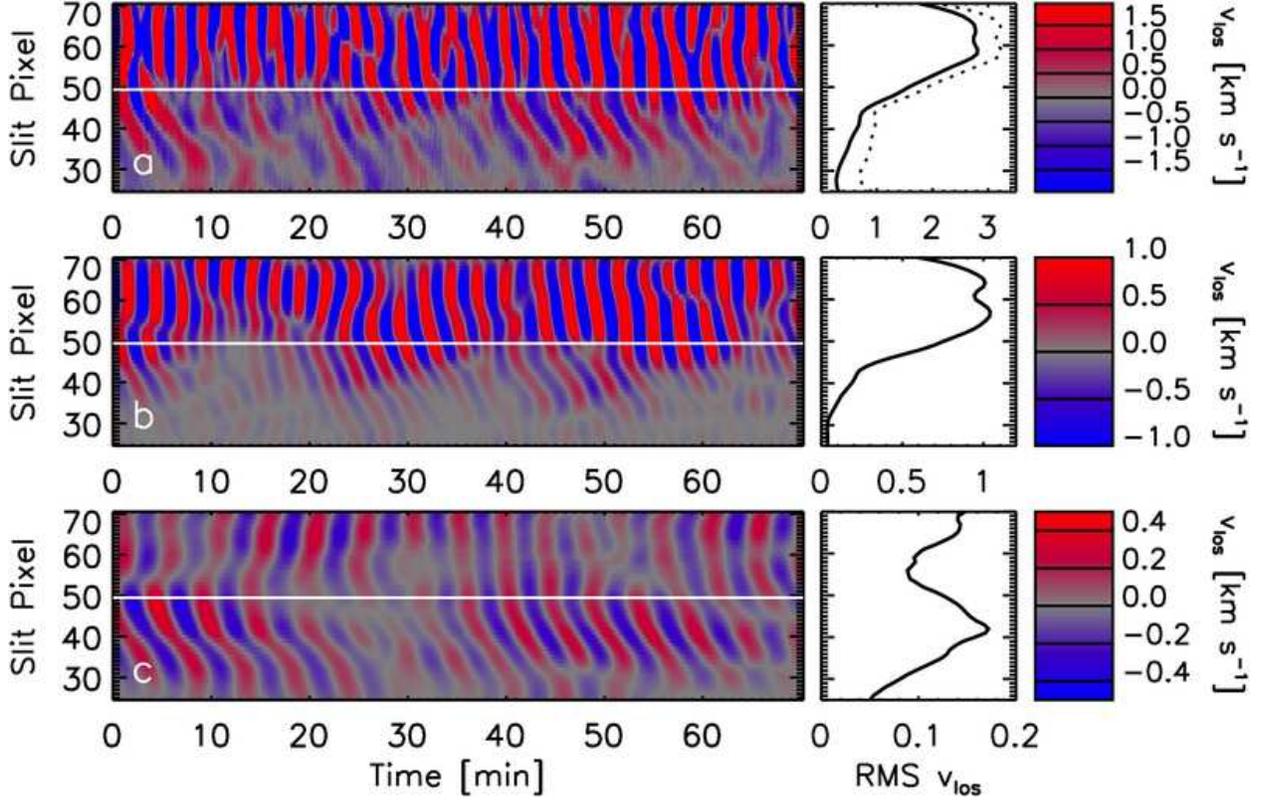}
\caption{Space-time plots covering the sunspot umbra and penumbra. {\emph{a}}) 
He\,{\sc{i}} velocities retrieved by the inversion. {\emph{b}}) He\,{\sc{i}} 
velocities after bandpass filtering in the range $2.5-3.5$~min. {\emph{c}}) 
He\,{\sc{i}} velocities after bandpass filtering in the range $4.5-5.5$~min. 
Values in panels {\emph{a}} and {\emph{b}} are clipped to enhance wavefront 
visibility within the penumbra. White lines mark the umbra/penumbra boundary, 
while right-hand panels show unclipped RMS LOS velocity. The dotted curve 
included in the upper RMS velocity panel shows the RMS velocity parallel to 
the field, after cosine correction for the inclination of the field from the 
observer's LOS.}
\label{fig:filt_5min}
\end{figure}

The termination of 3-min wave patterns at the umbra/penumbra boundary noted by 
\citet{2004A&A...424..671K} and \citet{2006SoPh..238..231K} and the fact that 
not all 3-min wavefronts can be traced out from the umbra into the penumbra 
has been used to suggest that RP waves are not associated with similar waves 
in the umbra. A simple check for the linkage of either 3-min or 5-min waves 
between the chromospheric umbra and penumbra can be made by bandpass filtering 
the He\,{\sc{i}} velocity time series. The spatial variation of He\,{\sc{i}} 
LOS velocities through the umbra and penumbra is presented in 
Fig.~\ref{fig:filt_5min} before and after filtering in the period ranges $2.5 
- 3.5$~min and $4.5 - 5.5$~min. It is clear that each of the 3-min umbral 
wavefronts has a rapidly diminishing counterpart in the penumbra 
(Fig.~\ref{fig:filt_5min}{\emph{b}}), while each of the 5-min (i.e., RP) 
wavefronts has an only somewhat weaker counterpart within the umbra 
(Fig.~\ref{fig:filt_5min}{\emph{c}}). These findings once again support the 
picture of a continuous variation of RP wave behaviour through the entirety of 
the sunspot atmosphere.

If RP waves are indeed the ``visual pattern'' of upward-propagating waves we 
expect that the wave velocity amplitude along the field would be essentially 
constant through the penumbra, because each wavefront will have experienced 
the same degree of wave growth caused by the decrease of density with 
altitude. Examination of the unfiltered He\,{\sc{i}} LOS velocity signal in 
Fig.~\ref{fig:filt_5min}{\emph{a}} shows that the RMS LOS velocity decreases 
throughout the penumbra. However, the correct quantity to consider is the 
RMS velocity along the magnetic field vector. This was obtained from the RMS 
LOS velocity (solid curve in the right-hand panel of 
Fig.~\ref{fig:filt_5min}{\emph{a}}) using the measured field inclinations from 
the observer's LOS -- resulting field-aligned RMS velocities are depicted in 
Fig.~\ref{fig:filt_5min}{\emph{a}} by a dotted curve. The field-aligned RMS 
velocities still show a decrease close to the umbra/penumbra boundary (from 
a greatly diminished contribution of transmitted power at 3-min period) but 
also a nearly constant value of $\sim$1~km~s$^{-1}$ over pixels 45 to 25 of 
the penumbra, lending more credence to the ``visual pattern'' scenario.

Our results support the conclusion of \citet{2006SoPh..238..231K} that 3-min 
umbral waves are not the source of 5-min RP waves. However, we have 
additionally shown that they are in fact different manifestations of the same 
form of wave generated by a common source at the photosphere, their 
differences arising from the transmitted wave power available for propagation 
along differently inclined field lines. As such, the observed behaviour of 
waves in both umbrae and penumbrae can be explained without the need for 
abrupt changes in either density or field orientation at the umbra/penumbra 
boundary as postulated by, e.g., \citet{2006A&A...456..689T, 
2007A&A...463.1153T}.

\section{CONCLUSIONS}
\label{sec:con}
We have provided evidence that velocity signatures of RP waves observed in the 
He\,{\sc{i}}\,10830~\AA\ multiplet are more compatible with upward-propagating 
waves than with trans-sunspot waves through careful consideration of the 
magnetic vector. Comparing the Fourier phase differences measured between 
paired pixels in the photosphere and chromosphere to the dispersion relation 
for field-aligned acoustic waves, modified for inclined fields, points toward 
such waves (i.e., essentially low-$\beta$ slow modes) being responsible for 
the visual pattern. 

Initially excited by a common source at the photosphere, waves experience 
increasing path length to the sampling height in the chromosphere with 
distance into the penumbra from travelling along increasingly inclined field 
lines -- a scenario previously suggested by, e.g., \citet{2003A&A...403..277R} 
and \citet{2006RSPTA.364..313B}. For essentially constant (or weakly 
increasing) propagation velocities, delays of increasing magnitude will be 
observed in the arrival times of wavefronts at increasing radial distance 
through sunspot penumbrae. It is the pattern of delayed wavefronts that gives 
rise to the apparent outward motion of RP waves which may also explain the 
large range of observed wave speeds -- the horizontal ``speed'' of the delayed 
wavefronts at the chromosphere depends on the rate at which the magnetic field 
inclines out through penumbrae, permitting either sub-sonic or super-sonic 
horizontal ``speeds'' for different magnetic geometries. This scenario also 
indicates that RP waves may occur at the edges of large pores since the 
existence of a penumbra is not necessary to support them; only sufficiently 
inclined field lines are required to direct the waves laterally.

\acknowledgments
The German Vacuum Tower Telescope is operated on Tenerife by the Kiepenheuer 
Insitute in the Spanish Observatorio del Teide of the Instituto de 
Astrof\'{i}sica de Canarias. The authors wish to extend their sincere thanks 
to R. Centeno, M. Collados, and J. Trujillo Bueno for providing this excellent 
data set for our analysis.

\newpage
\clearpage

\begin{table}
\begin{center}
\caption{Spatial pairings between Si\,{\sc{i}} and He\,{\sc{i}} pixels 
and their corresponding magnetic field inclinations in solar 
coordinates\label{tab:pixel_pairs}}
\begin{tabular}{ccccc}
\tableline
\tableline
\multicolumn{2}{c}{Si\,{\sc{i}}}	&  \multicolumn{2}{c}{He\,{\sc{i}}}	&  Notes$^{\dagger}$\\
Spatial	&  Inclination	&  Spatial	&  Inclination	&  \\
Pixel	&  (\degr)	&  Pixel	&  (\degr)	&  \\
\hline
56	&  172		&  55		&  165		&  a\\
55	&  170		&  54		&  165		&  a\\
54	&  169		&  53		&  164		&  b\\
53	&  167		&  52		&  164		&  b\\
52	&  165		&  51		&  163		&  c\\
51	&  163		&  50		&  163		&  c\\
50	&  161		&  49		&  163		&  c\\
49	&  160		&  48		&  163		&  d\\
48	&  158		&  47		&  163		&  d\\
47	&  157		&  46		&  162		&  d\\
47	&  157		&  45		&  161		&  e\\
46	&  155		&  44		&  160		&  e\\
45	&  153		&  43		&  158		&  e\\
44	&  151		&  42		&  156		&  f\\
43	&  150		&  41		&  153		&  f\\
42	&  148		&  40		&  150		&  f\\
41	&  146		&  39		&  147		&  g\\
40	&  145		&  38		&  145		&  g\\
39	&  143		&  37		&  143		&  g\\
39	&  143		&  36		&  142		&  h\\
38	&  141		&  35		&  139		&  h\\
37	&  139		&  34		&  138		&  h\\
36	&  136		&  33		&  136		&  h\\
35	&  134		&  32		&  135		&  h\\
\tableline
\end{tabular}
\tablerefs{$^{\dagger}$ Panel of Fig.~\ref{fig:fft_upw_pro} in which the 
resulting Fourier phase difference spectra are overplotted}
\end{center}
\end{table}

\end{document}